\documentclass[aps,nofootinbib,prd,preprint,preprintnumbers]{revtex4}
\usepackage{listings}
\usepackage{xspace}
\usepackage{fp}
\usepackage{ifthen}
\usepackage{ragged2e}
\usepackage{rotating}
\usepackage{graphicx,graphics}
\usepackage{amsmath}
\usepackage{textcomp}
\usepackage{amssymb}
\usepackage{titlesec}
\usepackage{color}
\usepackage{multirow}
\usepackage{amsfonts}
\usepackage{wasysym}
\DeclareMathAlphabet{\mathpzc}{OT1}{pzc}{m}{it}

\usepackage{hyperref}
\usepackage{color}

\def\21{$\mathrm{SU(2)_L \otimes U(1)_Y}$ }

\newcommand{\ESFM}{Departamento de F\'{\i}sica, Escuela Superior de 
F\'{\i}sica y Matem\'aticas del Instituto Polit\'ecnico Nacional \\ 
Unidad Adolfo L\'opez Mateos, Edificio 9, 07738 Ciudad de M\'exico, Mexico}

\newcommand{\Cinvestav}{Departamento de F\'{\i}sica, Centro de
  Investigaci{\'o}n y de Estudios Avanzados del IPN\\ Apdo. Postal
  14-740 07000 Ciudad de M\'exico, Mexico}
  
\begin{document}

\preprint{}

\title{K-essence and kinetic gravity
  braiding models in two-field measure theory}

\author{R. Cordero~$^1$}\email{cordero@esfm.ipn.mx}
\author{O. G. Miranda~$^2$}\email{omr@fis.cinvestav.mx}
\author{M. Serrano-Crivelli~$^1$} 
\affiliation{$^1$~\ESFM}
\affiliation{$^2$~\Cinvestav}

\begin{abstract}
We show that, in the context of the two-field measure
  theory, any k-essence model leads to the existence of a fluid made
  of non-relativistic matter and cosmological constant that would
  explain the dark matter and dark energy problem at the same time.
  On the other hand, kinetic gravity braiding models can lead to 
  different behaviors, such as phantom dark energy, stiff matter, and a
  cosmological constant. For stiff matter, there even exists
  the case where the scalar field does not have any effect in the
  dynamics of the Universe.
\end{abstract}


\maketitle
\section{Introduction}
The dark matter and dark energy problem are fundamental issues to be
solved in this century. The discovery of the accelerating expansion of
the Universe~\cite{Riess:1998cb,Perlmutter:1998np} is one of the most
intriguing and puzzling questions in cosmology
today~\cite{Sahni:2004ai,Alam:2004jy,Sahni:2006pa,Feng:2004ad,Durrer:2007re}. Many
proposals have been discussed in the literature, ranging from physics
beyond the Standard Model of elementary particles, such as
supersymmetry, modifications of the gravitational theory (MOND), the
cosmological constant, quintessence
field~\cite{Ratra:1987rm,Copeland:2006wr,Linder:2007wa}, brane
cosmology scenario~\cite{Deffayet:2001pu,Shtanov:2002mb} and k-essence
models~\cite{ArmendarizPicon:1999rj,ArmendarizPicon:2000dh,ArmendarizPicon:2000ah,Melchiorri:2002ux,Chiba:1999ka,Chiba:2002mw,Chimento:2003zf,Chimento:2003ta}. Among
the former proposals the cosmological constant is the simplest choice
although it needs a fine tuning value. There are different approaches
that try to model the late-time accelerated expansion of the Universe
including modified gravity at large distances via extradimensions or
including a dark energy component that does not fulfill the strong
energy condition~\cite{Amendola:2015ksp} or new proposals, as the case
of k-essence theories.

In this kind of models the scalar field plays an important role in
order to describe the dark energy problem.The scalar fields can
reproduce the dynamical effects of the cosmological constant at late
times and could have a very interesting and adequate behaviour at
early times.  Scalar fields models with non-canonical kinetic terms
have been proposed as an alternative to describe the dark energy
component of the Universe. The k-essence type Lagrangians were
introduced in several contexts, for example, as a possible model for
inflation~\cite{ArmendarizPicon:1999rj,Garriga:1999vw}. Later on
k-essence models were analyzed as an alternative to describe the
characteristics of dark energy and as a possible mechanism of unifying
dark energy and dark matter~\cite{Scherrer:2004au}. Purely kinetic
k-essence
models~\cite{dePutter:2007ny,Gao:2010ia,Cruz:2008cwa,Sur:2008tc} are,
in some sense, as simple as quintessence because they depend only on a
single function $F$ by means of the expression for the density
Lagrangian ${\cal{L}} = F(X)$, where the kinetic term is $X=
-g^{\mu\nu}\partial _\mu \phi \partial _\nu \phi$, $g_{\mu \nu}$ is
the metric of the spacetime, and $\phi$ is the scalar field.  Among
these fields generalized
galileons~\cite{Nicolis:2008in,deRham:2010eu,Goon:2010xh,Goon:2011qf}
have been studied in detail recently. They have the capacity to induce
an accelerated expansion of the Universe when no matter interaction is
present an additionally at short distances they possess the Vainshtein
screening mechanism. These scalar field models are second-order
derivative theories but with the appreciated property that their
equations of motions are second order.  A nice example of this kind of
models is the so-called kinetic gravity braiding models that are constructed by means
of a Lagrangian that contain a D'Alambertian operator and an arbitrary
function of a non-canonical kinetic terms \cite{Deffayet:2010qz}. The
kinetic gravity braiding models have the characteristic that the
energy density contains the Hubble parameter and the pressure depends
on the second derivatives of the scalar fields which give rise to very
interesting effects
\cite{Pujolas:2011he,Kimura:2010di,Kimura:2011td,Maity:2012dx}.

Recently, the Two Field Measure Theory (TMT) has also been proposed as
a possible tool in the search for an explanation to the observations
that leads to the dark energy and dark matter
hypothesis~\cite{Guendelman:2012gg,Ansoldi:2012pi}. In this theory, a
new measure is defined to build and action that has extra
contributions, additional to the usual measure of general relativity
($\sqrt{-g}$). 
Different actions have been studied in
  this
  context~\cite{Guendelman:2013ke,Guendelman:2015rea,Guendelman:2015jii}. 

In this work, we study both k-essence and kinetic gravity braiding models that could
lead to rich and interesting results in the framework of the TMT. We
will show that any k-essence model, in the TMT, will lead to dark
matter and dark energy in a natural way. On the other hand, the
kinetic gravity braiding models present a richer spectrum of possibilities. Even in
the simplest cases for these models can get either a cosmological
constant or stiff matter \cite{Zeldovich:1972zz,Chavanis:2014lra}.

The paper is organized as follows. In Section 2 the TMT is presented
where its motivation is enhanced. The next Section contains the
equations of motion for a general k-essence model in the context of
the TMT. Besides it is shown that any arbitrary k-essence model can
describe a dark matter and dark energy behavior in an unified way. In
Section 4, we apply the TMT to kinetic gravity braiding models, starting with a
discussion about the imperfect fluid form of these models and then, we
investigate what kind of fluid appears when a TMT contribution is
added. Finally we give our conclusions.

\section{The two field measure theory}
In order to discuss the k-essence theories in the context of the TMT, we will describe in this section the main
characteristic of the TMT, following closely the discussion given
in~\cite{Guendelman:2006wr}.

We will work in a Friedmann-Robertson-Walker metric 
\begin{equation}
\label{FRW1}
ds^{2}=g_{00}-a^{2}\left(t\right)\left[\frac{dr^{2}}{1-kr^{2}}+
r^{2}\left(d\theta^{2}+\sin^{2}\theta d\phi^{2}\right)\right], 
\end{equation}
where  $k=-1,0,1$, with the signature  $(+,-,-,-)$ \footnote{The signature in~\cite{Guendelman:2006wr} is $(-,+,+,+)$. }.

First it is important to consider an action that contains two
different terms: the usual metric, $g_{\mu\nu}$, and an additional
term with a new measure, $\Phi$, in the same space-time: 
\begin{equation}
\label{TMTACTION1}
S=\int \mathcal{L}_{1}\sqrt{-g}d^{4}x+\int \mathcal{L}_{2}\Phi d^{4}x.
\end{equation}
The new measure can be constructed, for instance, with four scalar
fields, $\varphi^{a}$ ($a=1,2,3,4$), 
\begin{equation}
\label{DEFMED1}
\Phi=\epsilon^{\mu\nu\alpha\beta}\epsilon_{ijkl}\partial_{\mu}\varphi^{i}\partial_{\nu}\varphi^{j}\partial_{\alpha}\varphi^{k}\partial_{\beta}\varphi^{l}.
\end{equation}
It is assumed that the Lagrangians in these equations are independent
from the fields $\varphi^{a}$. It is also assumed that all fields
are independent. Therefore, any relation among the fields will be a
consequence of the equation of motion. 

In particular, it can be shown that the variation of the action will
lead to the equation of motion~\cite{Guendelman:1998ms}
\begin{equation}
\label{TMTEQMOV1}
A^{\mu}_{i}\partial_{\mu}\mathcal{L}_{2}=0,
\end{equation}
with 
\begin{equation}
A^{\mu}_{i}=\epsilon^{\mu\nu\alpha\beta}\epsilon_{ijkl}\partial_{\nu}\varphi^{j}\partial_{\alpha}\varphi^{k}\partial_{\beta}\varphi^{l}, 
\end{equation}
and, since we have~\cite{Guendelman:1998ms} 
\begin{equation}
\label{TMTDET1}
\det\left(A^{\mu}_{i}\right)=\frac{4^{-4}}{4!}\Phi^{3}  
\end{equation}
an important restriction can be obtained for $\Phi \neq0$:
\begin{equation}
\label{TMTREST1}
\mathcal{L}_{2}=constant.
\end{equation}
 In this work, we will follow closely the work of
  Guendelman, \textit{et. al}~\cite{Guendelman:2012gg}, where they
  consider an action of the form
\begin{eqnarray}
\label{accion1}
\nonumber
S=\int\frac{R}{16\pi G}\sqrt{-g}d^{4}x+\int\frac{K\left(\phi,\nabla_{\mu} \phi\right)}{\sqrt{-g}}\Phi d^{4}x+\int K\left(\phi,\nabla_{\mu}\phi\right)d^{4}x,
\end{eqnarray}
where
\begin{align}
\label{K1}
K\left(\phi,\nabla_{\mu}\phi\right)=V\left(\phi\right)\sqrt{-g}\sqrt{1+g^{\mu\nu}\nabla_{\mu} \phi \nabla_{\nu} \phi} 
\end{align}
is a field of the type of DBI theories. 
In this case, the restriction given by Eq.~(\ref{TMTREST1}) turns out to be
\begin{align}
\label{RESGUE1}
\frac{K}{\sqrt{-g}}=V\left(\phi\right)\sqrt{1+\dot{\phi}^{2}}=M,
\end{align}
with $M$ a constant and $\phi$ is a function of time, $t$. With the
help of the energy-momentum tensor
\begin{equation}
\label{TENSOR1}
T_{\mu\nu}=\frac{2}{\sqrt{-g}}\frac{\delta S_{m}}{\delta g^{\mu\nu}},
\end{equation}
it is possible to obtain the energy density and
pressure of the fluid which are given by
\begin{eqnarray}
\label{GUERHO1}
\varepsilon&=&\left(1+\frac{\Phi}{\sqrt{-g}}\right)\frac{V\left(\phi\right)\dot{\phi}}{\sqrt{1+\dot{\phi}^{2}}} -V\left(\phi\right)\sqrt{1+\dot{\phi}^2}\\ \nonumber 
\label{GUEP1}
\mathcal{P}&=& -V\left(\phi\right)\sqrt{1+\dot{\phi}^2},
\end{eqnarray}
where we have used the definitions for energy density and
pressure~\cite{Pujolas:2011he} (given in Section 4):
\begin{align}
    \varepsilon = T_{\mu \nu}u^{\mu} u^{\nu}, \qquad \mathcal{P} =- \frac{1}{3}T^{\mu\nu} \perp_{\mu\nu}.
\end{align}
By solving the Lagrange equation for the Lagrangian density in Eq.~(\ref{accion1}) 
\begin{align}
\label{GUEEQMOV1}
-\frac{\partial}{\partial t}\left(\frac{V\left(\phi\right)\Phi \dot{\phi}}{\sqrt{1+\dot{\phi}^{2}}}+\frac{V\left(\phi\right)\sqrt{-g}\dot{\phi}}{\sqrt{1+\dot{\phi}^{2}}}\right)=\frac{\partial V}{\partial \phi}\left(\Phi\sqrt{1+\dot{\phi}^{2}}+\sqrt{-g}\sqrt{1+\dot{\phi}^{2}}\right),
\end{align} 
it is possible to find an expression for $\Phi$ in terms of 
$\phi$~\cite{Guendelman:2012gg}:
\begin{align}
\label{GUEPHI1}
\Phi=\frac{C\left(r,\theta\right)}{V^{2}-M^{2}}-\sqrt{-g},
\end{align}
where $C(r, \theta)$ is an integration constant given by 
\begin{align}
C\left(r, \theta\right)=\frac{Dr^{2}\sin\theta}{\sqrt{1-kr^{2}}}, 
\qquad D=constant.
\end{align}

Substituting $\Phi$ in the energy density and pressure gives
\begin{eqnarray}
\label{GUERHOFINAL1}
\varepsilon_{T}&=&M+\frac{D}{Ma^{3}}, \\ \nonumber \\
\label{GUEPFINAL1}
{\mathcal P}_{T}&=&-M. % \\ \nonumber \\
\end{eqnarray}
The parameter density is 
\begin{eqnarray}
\label{GUEPARFINAL1}
\omega&=&\frac{{\mathcal P}_{T}}{\varepsilon_{T}}=\frac{-1}{1+\frac{D}{M^{2}a^{3}}}.
\end{eqnarray}
The expression for $\varepsilon_T$ has two contributions: a constant term, $M$, and a dust-like term $a^{-3}$. It is shown that the inhomogeneous perturbations correspond to a particle fluid. This is achieved by computing the covariant derivative of the energy momentum tensor. The four velocity is defined as
\begin{align}
\label{fourvel}
u_{\mu}=\frac{\nabla_{\mu}\phi}{\sqrt{\nabla_{\alpha}\phi \nabla^{\alpha}\phi}}.
\end{align}
The energy momentum tensor has the perfect fluid form
\begin{align}
\label{tenpert}
T_{\mu\nu}=\varepsilon_{d}u_{\mu}u_{\nu}+Mg_{\mu\nu},
\end{align}
where
\begin{align}
\label{rhod}
\varepsilon_{d}=\left(1+\frac{\Phi}{\sqrt{-g}}\right)\frac{\left(\partial_{\alpha}\phi \partial^{\alpha}\phi\right)V\left(\phi\right)}{\sqrt{1+g^{\mu\nu}\partial_{\mu}\phi \partial_{\nu}\phi}}.
\end{align}
The covariant derivative of the second term in Eq.~(\ref{tenpert}) is
zero due to the presence of the metric tensor.  The covariant
derivative of the first term,
\begin{align}
\label{covder1}
\nabla_{\mu}\left(\varepsilon_{d}u^{\mu}u^{\nu}\right)=\nabla_{\mu}\left(\varepsilon_{d}u^{\mu}\right)u^{\nu}+\varepsilon_{d}u^{\mu}\nabla_{\mu}u^{\nu},
\end{align} 
is also zero due to the orthogonality property of the vectors $u^{\nu}$ and $u^{\mu}\nabla_{\mu}u^{\nu}$ and the conservation of the current $J^\mu = \varepsilon_d u^\mu$.
%%%%%%%%%%%%%%%%%%%%%%%%%%%%%%%%%%%%%%%%%%%%%%%%%%%%%
%}
\section{k-essence models and the two field measure theory}
Once we have introduced the TMT and the restriction on the new
Lagrangian, $L_2$, we turn our attention to the specific case of
k-essence models where we have the function
\begin{equation}
\label{GENLAG1}
K\left(\phi,\partial_{\mu}\phi\right)=G_2\left(\phi, X \right)\sqrt{-g},  
\end{equation}
that represents the more general k-essence
models~\cite{ArmendarizPicon:2000dh}, with $X=\nabla_{\alpha} \phi
\nabla^{\alpha} \phi/2$ being the kinetic term. In this case the
restriction on ${L}_{2}$ will be given as
\begin{align}
\label{GENREST1}
G_2\left(\phi, X\right)=M_{3},
\end{align}
with $M_{3}$ an arbitrary constant. We can therefore consider the 
following action in the TMT: 
\begin{eqnarray}
\label{GENACC1}
S_{m}=\int \left[G_2\left(\phi, X\right)\Phi+G_2\left(\phi, X\right)\sqrt{-g}\right]d^{4}x.
\end{eqnarray}
If we vary this action with respect to the metric, $g^{\mu\nu}$, and employ the 
usual expression for the energy momentum tensor given in  Eq.~(\ref{TENSOR1}), we obtain
\begin{align}
\label{GENTENSOR1}
T_{\mu\nu}=\left(1+\frac{\Phi}{\sqrt{-g}}\right)G_2,_{X} \nabla_{\mu} \phi \nabla_{\nu} \phi - g_{\mu\nu}G_2.  
\end{align}
Through this paper we are using the abbreviated notation,
$G,_{X}=dG/dX$ or $G,_{X}=\partial G/ \partial X$.

From these tensor, we can write down the total energy and pressure,
that includes the contribution from TMT, as
\begin{eqnarray}
\label{GENRHOTOT1}
\varepsilon_{T}&=&\left(1+\frac{\Phi}{\sqrt{-g}}\right) 2 XG_2,_{X} - G_2, \\ \nonumber \\
\label{GENPTOT1}
\mathcal{P}_{T}&=& -M_{3}.
\end{eqnarray}
We can find and expression for  $\Phi$ by using the Lagrange equations. For this purpose we use the following relations
\begin{eqnarray}
\frac{\partial \mathcal{L}_{m}}{\partial \phi}=G_2,_{\phi} \psi, \\ \nonumber \\
\frac{\partial \mathcal{L}_{m}}{\partial \dot{\phi}}=-G_2,_{X}\dot{\phi}\psi,
\end{eqnarray}
where  $\psi=\Phi+\sqrt{-g}$  and $\mathcal{L}_{m}$ is given by 
\begin{equation}
\label{GENLAGRANDEN1}
\mathcal{L}_{m}=G_2\left(\phi, X\right)\Phi+G_2\left(\phi, X\right)\sqrt{-g}.
\end{equation}
The Lagrange equations will be then expressed as 
\begin{align}
\label{GENLAGRAEC1}
\sqrt{2X}\frac{d}{d\phi}\left(\psi G_2,_{X}\sqrt{2X}\right)=G_2,_{\phi} \psi. 
\end{align}
If we use now the restriction~(\ref{GENREST1}) on $G_2$, we can find a relation between $G_2,_{X}$ and $G_2,_{\phi}$ : 
\begin{eqnarray}
dG_2=G_2,_{\phi}d\phi+G_2,_{X}dX,\\ \nonumber \\
\label{GENKPHI1}
G_2,_{\phi}=-G_2,_{X}\frac{dX}{d\phi}.
\end{eqnarray}
If we calculate the differential in the left hand side of equation~(\ref{GENLAGRAEC1})%
\begin{align}
d\left(\psi G_2,_{X}\sqrt{2X}\right)=G_2,_{X}\sqrt{2X}d\psi+\psi \left(G_2,_{XX}\sqrt{2X}+\frac{G_2,_{X}}{\sqrt{2X}}\right)dX, 
\end{align}
we can substitute this expression and Eq.~(\ref{GENKPHI1}) 
in the Eq.~(\ref{GENLAGRAEC1}) and find that 
\begin{equation}
\frac{d \psi}{\psi}=\left(-\frac{1}{X}-\frac{G_2,_{XX}}{G_2,_{X}}\right)dX ,
\end{equation}
which can be solve to give
\begin{equation}
\label{GENPSI1}
\psi=\frac{C_{3}\left(r,\theta\right)}{XG_2,_{X}}.
\end{equation}
From this last expression for  $\psi$ we obtain $\Phi$
\begin{align}
\label{GENPHI1}
\Phi=\frac{C_{3}\left(r,\theta\right)}{XG_2,_{X}}-\sqrt{-g}.
\end{align}
Finally, by adopting a particular expression for the constant $C_{3}$ we cancel out the dependence on the energy density in the spatial variables,
\begin{align}
C_{3}\left(r,\theta\right)=\frac{D_{3}r^{2}\sin \theta}{\sqrt{1-kr^{2}}}, 
\end{align}
where $D_{3}$ is a constant.
After this last step, the total energy density and the total pressure
will be written as
\begin{eqnarray}
\label{GENRHOFINAL1}
\varepsilon_{T}&=&\frac{D_{3}}{a^{3}} - M_{3}, \\ \nonumber \\
\label{GENPFINAL1}
\mathcal{P}_{T}&=& M_{3}.
\end{eqnarray}
As in the previous section, for this case the inhomogeneous perturbations also correspond to dust particles. The energy momentum tensor, Eq.~(\ref{GENTENSOR1}), can be expressed as 
\begin{align}
\label{tenpertk}
T_{\mu\nu}=\varepsilon_{d}u_{\mu}u_{\nu} - Mg_{\mu\nu},
\end{align}
where $u_{\mu}$ is given by Eq.~(\ref{fourvel}) and $\varepsilon_{d}$ is now
\begin{align}
\varepsilon_{d}=\left(1+\frac{\Phi}{\sqrt{-g}}\right)\left(\nabla_{\alpha}\phi \nabla^{\alpha} \phi \right)G_{2,X}.
\end{align}
The expression in Eq.~(\ref{tenpertk}) is exactly the same as Eq.~
(\ref{tenpert}). The only change is in the
expression for $\varepsilon_{d}$. Therefore, the
covariant derivative of the tensor (\ref{GENTENSOR1}) is zero, as it
is expected.  \\ 
We can see that we have obtain a general result for
k-essence theories in the TMT. In this formulation, we will
always obtain a dependence $a^{-3}$ for the energy density and a
cosmological constant therefore it is possible to model dark energy
and dark matter. This is a general result for any k-essence action
and, therefore, it applies for particular k-essence action as, for
example, linear actions
\begin{equation}
\label{POLAC1}
K\left(\phi,\partial_{\mu} \phi\right)=V\left(\phi\right)\left(1+AX\right),
\end{equation}
where  $A$ is a constant and
$X=\frac{1}{2}g^{\mu\nu}\partial_{\mu} \phi \partial_{\nu} \phi$; or 
logarithmic actions 
\begin{equation}
\label{LOGAC1}
K\left(\phi, \partial \phi\right)=V\left(\phi\right)\frac{1}{b}\ln\left(\frac{c^{\star}X^{\beta}}{1-\alpha^{\star}X^{\beta}}\right),
\end{equation}
where  $\alpha^{\star}$, $c^{\star}$, $b$ and $\beta$ are constants. 

%%%%%%%%%%%%%%%%%%%%%%%%%%%%%%%%%%%%%%%%%%%%%%%%%%%%%

\section{Kinetic gravity braiding models and the TMT}
As we have seen, the TMT makes that k-essence action will be able to
model both dark matter and dark energy. In this section, we pay
attention to a different case, where we consider the kinetic gravity braiding models,
and we will show that its behaviour in the TMT can be more complex. 

It is well known that the more general scalar field theories in four
dimensions that have second order (or less) equations of motion are
described by the Lagrangian~\cite{Deffayet:2011gz}
\begin{align}
\mathcal{L}=\sum_{i=2}^{5}\mathcal{L}_{i},
\end{align}
where 
\begin{eqnarray}
\mathcal{L}_{2}&=&K\left(\phi,X\right),\\ \nonumber\\
\mathcal{L}_{3}&=&-G_{3}\left(\phi,X\right)\square\phi, \\ \nonumber \\
\mathcal{L}_{4}&=&G_{4}\left(\phi,X\right)R+G_{4},_{X}\left[\left(\square \phi\right)^{2}-\left(\nabla_{\mu}\nabla_{\nu}\phi\right)\left(\nabla^{\mu}\nabla^{\nu}\phi\right)\right],\\ \nonumber \\
\mathcal{L}_{5}&=&G_{5}\left(\phi,X\right)G_{\mu\nu}\left(\nabla^{\mu}\nabla^{\nu}\phi\right)-\frac{1}{6}G_{5},_{X}\left[\left(\square \phi\right)^{3}-3\left(\square\phi \right)\right. \\ \nonumber
&&\left.  \left(\nabla_{\mu}\nabla_{\nu}\phi\right)\left(\nabla^{\mu}\nabla^{\nu}\phi\right)+2\left(\nabla^{\mu}\nabla_{\alpha}\phi\right)\left(\nabla^{\alpha}\nabla_{\beta}\phi\right)\left(\nabla^{\beta}\nabla_{\mu}\phi\right)\right]. 
\end{eqnarray} 
Here $G_{i},_{X}\equiv \partial G_{i}/ \partial X$, $R$ is the Ricci scalar, and  $G_{\mu\nu}$ is the  Einstein tensor.

We will consider now the kinetic gravity braiding action $\mathcal{L}_{3}$
\begin{align}
\label{BRAAC1}
K\left(\phi,X, \ddot{\phi}\right)=-G_{3}\left(\phi,X\right)\square \phi \sqrt{-g}.
\end{align}

In order to have a better understanding of the kinetic gravity braiding case, we will
discuss first the hydrodynamic fluid picture and show how the kinetic gravity braiding results 
can be generalized for the TMT case. 

%%%%%%%%%%%%%%%%%NEW
\subsection{The imperfect fluid in kinetic gravity braiding models and the TMT theory.}

We can start our discussion by considering first the kinetic gravity braiding action without 
TMT: 
\begin{align}
\label{BR1}
S_{\phi}=\int d^{4}x\sqrt{-g}\quad \square\; \phi G\left(X,\phi\right),
\end{align}
where we can write $X$ and $\square\phi$ in covariant form, that is,
$X=\nabla^{\mu}\phi\nabla_{\mu}\phi/2$, $\square \;
\phi=\nabla^{\mu}\nabla_{\mu}\phi$
and the function $G$ is arbitrary. 

From the computations in Ref.~\cite{Deffayet:2010qz}, the equation of
motion for the field $\phi$ is
%%%%%%%%%%%%%%%%%%%%%%%%%%%%%%%equation of motion
\begin{eqnarray}
\label{BR2}
    \nabla_{\mu}J^{\mu} &=& \mathcal{P}_{\phi}, \\ \nonumber
\end{eqnarray}
where %$J_{\mu}$ is a Noether current and
%%%%%%%%%%%%%%%%%%%%%%Noether
\begin{eqnarray}
    \label{BR3}
J_{\mu} &=& \left(\mathcal{L}_{X}-2G_{\phi}\right)\nabla_{\mu}\phi-G_{X}\nabla_{\mu}X, \\ \nonumber \\
    \label{BR4}
\mathcal{P}_{\phi} &=& - \nabla^{\lambda}\phi \nabla_{\lambda} G_{\phi}.
\end{eqnarray}
The corresponding energy-momentum tensor, $T_{\mu\nu}$, is given by
%%%%%%%%%%%%%%%%%%% energy momentum
\begin{align}
    \label{BR5}
    T_{\mu\nu} = \frac{2}{\sqrt{-g}}\frac{\delta S_{\phi}}{\delta g^{\mu\nu}} = \mathcal{L}_{X}\nabla_{\mu} \phi \nabla_{\nu} \phi - g_{\mu\nu}\mathcal{P} - \nabla_{\mu} G \nabla_{\nu} \phi - \nabla_{\nu} G \nabla_{\mu} \phi.
\end{align}

In Ref.~\cite{Pujolas:2011he}, that we follow
closely in this subsection, the authors rewrite these expressions using the hydrodynamic fluid picture. 
%%%%%%%%%%%%%%%%%%%%%%%%%%%%%%%%%%%%%%%%%%%%%%%%%%%%%%%% MAybe write something about fluid picture
Let's define some useful quantities used to describe relativistic fluids.
We set a local rest frame defining the four-velocity, $u_{\mu}$,
%%%%FOUR VELOCITY
\begin{align}
    \label{BR21}
    u_{\mu} \equiv \frac{\nabla_{\mu}\phi}{\sqrt{2x}},
\end{align}
which is normalized, $u_{\mu}u^{\mu}=1$. The four-acceleration is
%%%%%%%%%%%%%%%%%%%%ACCELERATION
\begin{align}
    \label{BR22}
    a_{\mu} \equiv \dot{u_{\mu}} \equiv  u^{\nu} \nabla_{\nu}u_{\mu},
\end{align}
where the acceleration is orthogonal to the velocity, $u_{\mu} a^{\mu}
= 0$, and the dot represents the material derivative along
$u^{\lambda}$,
%%%%%%%%%MATERIAL DERIVATIVE
\begin{align}
    \label{BR23}
    \dot{\left(  \; \right)} = \frac{d}{d \tau} \left( \; \right) = u^{\lambda} \nabla_{\lambda} \left( \; \right).
\end{align}
The expansion, $\theta$, and the diffusivity, $\kappa$, are defined
\begin{eqnarray}
    \label{BR24}
    \theta = \nabla_{\mu} u^{\mu}, \quad  \kappa = 2 X G_{X}.
\end{eqnarray}
Under these scheme, the current in Eq.~(\ref{BR3}) is written as 
%%Noether in hydro
\begin{align}
    \label{BR6}
    J_{\mu} = \left(-2 \dot{\phi} G_{\phi} + \kappa\theta\right) u_{\mu} - \kappa a_{\mu}.
\end{align}
When dissipation is modeled, the usual condition that diffusion $k^\mu$ must
satisfy is $u_{\mu} \kappa^{\mu} = 0$, that is, the diffusion is only
spatial~\cite{Andersson2007}. The last term in Eq.~ (\ref{BR6})
corresponds to the diffusion, say $\kappa_{\mu} = -\kappa a_{\mu}$
(note that $u_{\mu} \kappa^{\mu} = 0$ is satisfied through $u_{\mu}
a^{\mu} = 0$). The equation of motion, Eq.~(\ref{BR2}), takes a nice
form if one use the density of charge, $n$, defined as
%%%%%%charge n
\begin{align}
    \label{BR7}
    n \equiv u^{\mu} J_{\mu} = -2 \dot{\phi} G_{\phi} + \kappa \theta. 
\end{align}
In the case, the aforementioned equation for $\phi$, Eq.~(\ref{BR2}), becomes
%%%%%%%%%%%%%%%%%% motion hydro
\begin{align}
    \label{BR8}
    \dot{n} + \theta n - \nabla_{\mu} \left(\kappa a^{\mu}\right) = \mathcal{P}_{\phi}.
\end{align}
Expressing the energy-momentum tensor requires of the energy density, $\varepsilon$, and the total isotropic pressure, $\mathcal{P}$,
%%%%%% energy density pressure
\begin{eqnarray}
    \label{BR9}
    \varepsilon & \equiv & T_{\mu\nu} u^{\mu} u^{\nu} = -2 X G_{\phi} + \theta m \kappa, \\ \nonumber \\
    \label{BR10}
    \mathcal{P} & \equiv & -\frac{1}{3}T^{\mu\nu} \perp_{\mu\nu} = -2 X G_{\phi} - \kappa \dot{m}.
\end{eqnarray}
The energy-momentum is
%%%%%%energy momentum tensor hydro
\begin{align}
    \label{BR11}
    T_{\mu\nu} = \varepsilon u_{\mu} u_{\nu} - \perp_{\mu\nu} \mathcal{P} + u_{\mu} q_{\nu} + u_{\nu} q_{\mu},
\end{align}
where $m = \sqrt{2X} = \dot{\phi}$ is the chemical potential and $q_{\mu} = -m \kappa a_{\mu}$ is the heat flux that, again, must be purely spatial ($u_{\mu} q^{\mu} = 0$).
Finally, the conservation of $T_{\mu\nu}$ leads to
%%%%%%%%%%%%%%%%%%%% energy momentum conservation
\begin{align}
    \label{BR12}
    u_{\nu} \nabla_{\mu} T^{\mu\nu} = \dot{\varepsilon} + \theta \left( \varepsilon + \mathcal{P} \right) - \nabla_{\lambda} \left( m \kappa a^{\lambda} \right) + m \kappa a_{\lambda}a^{\lambda} = 0.
\end{align}

Up to this point, the action is such that the only contribution comes
from the usual measure $\sqrt{-g} \; d^{4}x$. Now we want to include
the contribution from TMT:
%%%%TMT action
\begin{align}
    \label{BR13}
    S=\int d^{4}x\sqrt{-g}\; \square\; \phi G\left(X,\phi\right) + \int d^{4}x\; \Phi \; \square\; \phi G\left(X,\phi\right). 
\end{align}
Notice the replacement of $\sqrt{-g}$ by $\Phi$ in the second integral of this last expression. The new equation of motion becomes:
%%%%TMT eq of motion
\begin{align}
    \label{BR14}
    \nabla_{\mu}\left[ 2 G_{\phi} \nabla^{\mu} \phi  - \square \phi G_{X} \nabla^{\mu}\phi + G_{X}\nabla^{\mu}X \right] + \\ \nonumber
    \frac{\Phi}{\sqrt{-g}}\nabla_{\mu}\left[ 2 G_{\phi} \nabla^{\mu} \phi  - \square \phi G_{X} \nabla^{\mu}\phi + G_{X}\nabla^{\mu}X \right] + \\ \nonumber
    \frac{1}{\sqrt{-g}} \nabla_{\mu} \left( G \nabla^{\mu} \Phi \right) = \nabla^{\lambda} \phi \nabla_{\lambda} G_{\phi} + \frac{\Phi}{\sqrt{-g}} \nabla^{\lambda} \phi \nabla_{\lambda} G_{\phi} + \square \phi G_{X} \nabla^{\lambda} \phi \frac{\nabla_{\lambda} \Phi}{\sqrt{-g}} - \frac{\nabla^{\lambda} \Phi}{\sqrt{-g}} \nabla_{\lambda}G.  \\ \nonumber 
\end{align}
Making use of the expressions in Eqs.~(\ref{BR3}) and~(\ref{BR4}),
this equation can be reduced to
%%%%%TMT eq of motion
\begin{align}
    \label{BR15}
    \nabla_{\mu}\left[ \left( \sqrt{-g} + \Phi \right) J^{\mu} \right]  + \nabla_{\mu} \left( G \nabla^{\mu} \Phi \right) = \mathcal{P}_{\phi} \left( \sqrt{-g}+ \Phi \right) + G_{\phi} \nabla^{\lambda} \nabla_{\lambda} \Phi.
\end{align}

The energy-momentum tensor can be expressed similar to equation (\ref{BR11}):
%%%%%%%%%%%%%%%%%%%%%%%% Energy momentum tensor tmt
\begin{align}
    \label{BR16} \nonumber
    T_{\mu\nu} = \left[ \varepsilon + \varepsilon \frac{\Phi}{\sqrt{-g}} + \sqrt{2X} G u_{\alpha} \frac{\nabla^{\alpha} \Phi}{\sqrt{-g}} + G \square \phi \frac{\Phi}{\sqrt{-g}} - 2 \sqrt{2X} G u^{\lambda}\frac{\nabla_{\lambda}\Phi}{\sqrt{-g}} \right] u_{\mu}u_{\nu} + 
    \\ \nonumber \\ \nonumber
    -\perp_{\mu\nu} \left[ \mathcal{P} + \mathcal{P} \frac{\Phi}{\sqrt{-g}} - \sqrt{2X} G u_{\alpha} \frac{\nabla^{\alpha} \Phi}{\sqrt{-g}} -  G \square \phi \frac{\Phi}{\sqrt{-g}} \right] \\ \nonumber \\ \nonumber
    + \left[ q_{\nu} + q_{\nu}  \frac{\Phi}{\sqrt{-g}} - \sqrt{2X} G
\perp^{\lambda}_{\nu} \frac{\nabla_{\lambda}\Phi}{\sqrt{-g}} \right] u_{\mu} + \left[ q_{\mu} + q_{\mu}  \frac{\Phi}{\sqrt{-g}} - \sqrt{2X} G
\perp^{\lambda}_{\mu} \frac{\nabla_{\lambda}\Phi}{\sqrt{-g}} \right] u_{\nu}. \\
\end{align}
From this tensor, the energy density, $\varepsilon_{T}$ , the pressure, $\mathcal{P}_{T}$ and the energy flow, $Q_{\nu}$, are:
%%%%%%%%%%%%%%%%%%%%%%%%%%%NEW DEF
\begin{eqnarray}
    \label{BR17}
    \varepsilon_{T} &=& \varepsilon + \varepsilon \frac{\Phi}{\sqrt{-g}} + \sqrt{2X} G u_{\alpha} \frac{\nabla^{\alpha} \Phi}{\sqrt{-g}} + G \square \phi \frac{\Phi}{\sqrt{-g}} - 2 \sqrt{2X} G u^{\lambda}\frac{\nabla_{\lambda}\Phi}{\sqrt{-g}}, \\ \nonumber \\
    \label{BR18}
    \mathcal{P}_{T} &=& \mathcal{P} + \mathcal{P} \frac{\Phi}{\sqrt{-g}} - \sqrt{2X} G u_{\alpha} \frac{\nabla^{\alpha} \Phi}{\sqrt{-g}} -  G \square \phi \frac{\Phi}{\sqrt{-g}}, \\ \nonumber \\
    \label{BR19}
    Q_{\nu} &=&  q_{\nu} + q_{\nu}  \frac{\Phi}{\sqrt{-g}} - \sqrt{2X} G
\perp^{\lambda}_{\nu} \frac{\nabla_{\lambda}\Phi}{\sqrt{-g}} .
\end{eqnarray}
These expressions contain the energy density, $\varepsilon$, and pressure, $\mathcal{P}$, of the kinetic gravity braiding models (Lagrangian \ref{BR1}) given by Eqs. (\ref{BR9}, \ref{BR10}), respectively. From the definition of $Q_{\nu}$, the condition $u_{\nu}Q^{\nu}=0$ is satisfied. The expression for the conservation of the energy-momentum tensor can be written as Eq.~(\ref{BR12}):
%%%%%CONSERVATION
\begin{align}
    \label{BR20}
     u_{\nu} \nabla_{\mu} T^{\mu\nu} = \dot{\varepsilon_{T}} + \theta \left( \varepsilon_{T} + \mathcal{P}_{T} \right) + \nabla_{\mu}Q^{\mu} - Q^{\nu}a_{\nu} + u_{\nu} Q^{\mu} \nabla_{\mu} u^{\nu}= 0.
\end{align}

%%equation%%%%%
\section{Some solutions to the equation of motion in cosmology}
In order to describe our Universe, we must take into account the
cosmological principle. This tell us that the field $\phi$ is only
time-depend.  An equivalent form of the equation of motion,
Eq.~(\ref{BR15}), is obtained by means of Lagrange equations:
\begin{align}
    \label{BR33}
    \frac{d^{2}}{dt^{2}}\left[\frac{\partial \mathcal{L}}{\partial \ddot{\phi}} \right] - \frac{d}{dt} \left[ \frac{\partial \mathcal{L}}{\partial \dot{\phi}} \right] + \frac{\partial \mathcal{L}}{\partial \phi} = 0.
\end{align}
From this point of view, the equation for $\phi$, not written in a covariant way, is:
\begin{eqnarray}
    \label{BR34}
    G_{\phi} \left( \ddot{\phi} + 3H \dot{\phi} \right) \left[ \sqrt{-g} + \Phi \right] - \frac{d}{dt} \left[ \left( 3 H \dot{\phi}^{2} G_{X} - 2 G_{\phi} \dot{\phi} + 3HG + G_{\phi} \dot{\phi}\right) \left[ \sqrt{-g} + \Phi \right] \right] \\ \nonumber \\ \nonumber
    + \frac{d}{dt}\left( G \frac{d}{dt}\left[ \sqrt{-g} + \Phi \right] \right) = 0.
\end{eqnarray}
Now that we have the equation of motion for the field $\phi$,
Eq.~(\ref{BR15}), and the expressions for the energy density and
pressure, Eqs.~(\ref{BR17},\ref{BR18}), respectively, we want to
calculate an expression for the new measure field, $\Phi$, so that we
can find the behaviour of the fluid represented by the above
equations. The approach is to consider special cases for the new measure $\Phi$ and the function
$G$ in the kinetic gravity braiding action Eq.~(\ref{BR13}) and try to solve the
differential equation (\ref{BR33}). We will consider the following
cases $G=cte$ and $G= G(X)$.  From the definition of
four velocity, Eq.~(\ref{fourvel}), the only non-zero component is:
\begin{align}
    \label{BR25}
    u_{0}=\dot{\phi}.
\end{align}
The expansion coefficient, $\theta$, becomes
\begin{align}
    \label{BR26}
    \theta = \nabla_{0}u^{0} = \partial_{0} u^{0} + \Gamma^{0}_{\mu 0} u^{0} = 3H.
\end{align}
An isotropic and homogeneous Universe is modeled via a perfect fluid,
meaning that the energy flow $Q_{\nu}$ vanish for all $\nu$, as can be
verified from Eq.~(\ref{BR19}). The energy density and pressure,
Eqs.~(\ref{BR17}, \ref{BR18}), take the structure:
\begin{eqnarray}
    \label{BR35}
    \varepsilon_T &=& \varepsilon \left( 1+ \frac{\Phi}{\sqrt{-g}} \right) - 
\frac{\dot{\phi} G }{\sqrt{-g}} a^{3} \frac{d}{dt}\left[ \frac{\Psi}{a^{3}} 
\right] + G \Box \phi \frac{\Phi}{\sqrt{-g}} , \\ \nonumber \\
    \label{BR36}
    \mathcal{P}_{T} &=& \mathcal{P} \left( 1+ \frac{\Phi}{\sqrt{-g}} \right) - 
\frac{\dot{\phi} G }{\sqrt{-g}} a^{3} \frac{d}{dt}\left[ \frac{\Psi}{a^{3}} 
\right] - G \Box \phi \frac{\Phi}{\sqrt{-g}},
\end{eqnarray}
where we have computed the covariant derivatives using the fact that
$\Psi = \sqrt{-g} + \Phi$ is a scalar density of weight one. If we
take the equation of motion, Eq.~(\ref{BR34}), together with the
restriction (\ref{TMTREST1}) in the form
 \begin{align}
     \label{BR53}
      G \Box \phi = B,
 \end{align}
 where $B$ is a constant, we can
 check the validity of the continuity equation,
 \begin{align}
     \label{BR54}
     \dot{\varepsilon_{T}} + 3H \left( \varepsilon_{T} + \mathcal{P}_{T} \right) = 0.
    \end{align}
The next step is to find the form of the new measure, $\Phi$, to give
the complete description of the fluid. We have to solve the system of
two equations, the equation of motion and the restriction, with three
unknowns: the new measure contained in $\Psi$ , the scale factor $a$
and the field $\phi$. So we need an extra equation. When we just have
a k-essence action, there is no appearance of the scale factor, $a$,
or of the Hubble factor, $H = \dot{a} /a$, but in the kinetic gravity braiding action,
the presence of $H$ is behind the D'Alembertian operator acting on
$\phi$. We know that the dynamics of the Universe is governed by
Friedmann equations so, the fluid represented by $\varepsilon_{T}$ and
$\mathcal{P}_{T}$ must obey:
 \begin{eqnarray}
     \label{BR46}
     \left( \frac{\dot{a}}{a} \right) ^{2} &=& \frac{8 \pi G}{3} \varepsilon_{T} - \frac{k}{a^{2}},
     \\ \nonumber \\ 
     \frac{\ddot{a}}{a} &=& - \frac{4 \pi G}{3} \left( \varepsilon_{T} + \mathcal{P}_{T} \right).
 \end{eqnarray}
 Here, $G$ is the Newton's constant and $k$ is the spatial curvature.
 %%%%%%%%%%%%%%%%%%%%%%%%%%%%%%%%%%%%%%%%%%%%%%%%%%%%%%%%%%%%%%%%%
 
\subsection{Stiff matter} 

The equation of motion, Eq. (\ref{BR34}) can be written in the following way
\begin{equation}
 \Psi \frac{G_\phi}{G} B + \frac{d}{dt} \left[ a^3 \frac{d}{dt} \left( \frac{G\Psi}{a^3} \right) - \Psi \frac{G_X}{G} \dot{\phi} B \right] = 0
\end{equation}
with the help of the restriction (\ref{BR53}). If we set $B=0$ the solution of the equation of motion can be obtained from the simplified relation
\begin{equation}
\label{simplified}
 \frac{d}{dt}\left(\frac{G\Psi}{a^3}\right) = \frac{A}{a^3},
\end{equation}
where $A$ is an integration constant. Considering the following form of the restriction
\begin{equation}
 \label{restriction}
    \nonumber
    G \Box \phi = G \left( \ddot{\phi} + 3H \dot{\phi} \right) = 
    G\frac{1}{a^{3}} \frac{d}{dt} \left( a^{3} \dot{\phi}\right) = 0,
\end{equation}
then its solution can be cast as $\dot{\phi} = C/a^{3}$ where $C$ is a constant. Taking into account this solution the expressions for the energy density can also be simplified:
\begin{equation}
 \label{BR41}
 \varepsilon_T = {\cal P}_T = A\frac{\dot{\phi} }{\sqrt{-g}},
\end{equation}
which turns out to be the equation of state for stiff matter \cite{Zeldovich:1972zz,Chavanis:2014lra}.

For a complete description of the model, we need to solve the equation
for the new measure $\Phi$, Eq.~(\ref{simplified}). However, if we use
Friedmann equation, Eq.~(\ref{BR46}), we can obtain the time
dependence of the energy density and pressure, Eq.~(\ref{BR41}). We
consider a flat Universe, $k=0$,
\begin{equation}
\left( \frac{\dot{a}}{a} \right)^{2} = -\frac{FA_{stiff}}{a^{6}},
\end{equation}
where we set $\sqrt{-g} = \alpha a^{3}$, $A_{stiff} = A C /\alpha$, $F = 8 \pi G/3$ and $\alpha$ is a constant. The solution of the Friedmann equation is
\begin{equation}
\label{BR56}
\frac{a^{3}}{3} = \sqrt{-FA_{stiff}} \left( t - t_{0} \right) + \frac{a_{0}^{3}}{3}.
\end{equation}
The energy density and pressure as a functions of time are:
 \begin{align}
     \label{BR57}
     \varepsilon_{T} = \mathcal{P}_{T} = -\frac{A_{stiff}}{9} \left[ \sqrt{-FA_{stiff}} \left( t - t_{0} \right) + \frac{a_{0}^{3}}{3} \right]^{-2}.
 \end{align}
It is very interesting to note that when $A=0$ and, therefore, $\Psi = D \frac{a^3}{G}$ (where $D$ is a constant) there is not any effect of the braided fields in the dynamics of the Universe because
$\varepsilon_T =0$.
\subsection{Function G is constant}

The equation of motion, Eq. (\ref{BR34}), forms a total derivative and
can be simplified:
\begin{eqnarray}
    \label{BR37}
    \nonumber
    \frac{d}{dt}\left[ G \frac{d}{dt} \Psi - 3 H G \Psi \right] &=& 0, \\ \nonumber \\     G a^{3} \frac{d}{dt} \frac{\Psi}{a^{3}} &=& A.
\end{eqnarray}
where $A$ is a real number. Another equation that must fulfill the fields comes from the restriction of TMT models
\begin{equation}
    \label{BR38}
    \nonumber
    G \Box \phi = G \left( \ddot{\phi} + 3H \dot{\phi} \right) =  
    G\frac{1}{a^{3}} \frac{d}{dt} \left( a^{3} \dot{\phi} \right) = B,
\end{equation}
where $B$ is a constant. The energy density and pressure can also be simplified:
\begin{eqnarray}
    \label{BR39}
    \varepsilon_{T} &=& -A\frac{\dot{\phi} }{\sqrt{-g}} + B\frac{\Phi}{\sqrt{-g}}, \\ \nonumber \\
    \label{BR40}
    \mathcal{P}_{T} &=& -A\frac{\dot{\phi}}{\sqrt{-g}} - B\frac{\Phi}{\sqrt{-g}}.
\end{eqnarray}
The terms $\varepsilon$ and $\mathcal{P}$  coming from the contribution of the kinetic gravity braiding model are zero, Eqs.~(\ref{BR9}, \ref{BR10}), because their dependence on the derivatives of $G$.
\begin{itemize}
\item \textbf{Cosmological constant:} If we select $A = 0$ and $B \neq 0$, we get
    \begin{align}
        \label{BR42}
        \varepsilon_{T} = - \mathcal{P}_{T} = B \frac{\Phi}{\sqrt{-g}}.
    \end{align}  
From the equation of motion, Eq.~(\ref{BR37}), the new measure $\Phi =
D a^{3}$ ($D$ is a constant coming from the integration). In this
case, the energy density and pressure are constants $\varepsilon_{T} =
- \mathcal{P}_{T} = BD/\alpha$. As in the stiff matter case, we can
use Friedmann equation for a flat Universe to obtain the expected
exponential expansion of Universe
    \begin{equation}
    \frac{a}{a_{0}} = \exp \left[ \sqrt{F A_{CC}} \left( t - t_{0} \right) \right],
    \end{equation}
    where $A_{CC} = BD/\alpha$. 
    \end{itemize}
In the general case when both, $A$ and $B$, are non zero we could not
obtain analytical expressions for $a, \varepsilon_T, \mathcal{P}_{T}$.
  
It is interesting to note that a general cosmological constant
behavior of the Universe can be obtained if it is chosen the new
measure as $\Phi = -\sqrt{-g}$ and therefore $\Psi =0$. The
Eq. (\ref{BR34}) for the field $\phi$ is trivially satisfied and the
expressions for the energy density and the pressure are the same as in
Eq. (\ref{BR42}).
 %%%%%%%%%%%%%%%%%%%%%%%%%%%%%%%%%%%%%%%%%%%%%%%%%%%%%%%%%%%%%%%%%%%%%%%%%%%%%%%%%%%%%%% 
\subsection{Function G depends only on $X$.}
  The energy density and pressure are represented by Eqs.~(\ref{BR35}, \ref{BR36}). The contributions of the kinetic gravity braiding model, Eqs.~(\ref{BR9}, \ref{BR10}), becomes:
  \begin{eqnarray}
      \label{BR43}
      \varepsilon &=& 6H\dot{\phi} X G_{X}, \\ \nonumber \\
      \mathcal{P} &=& -2 X G_{X} \ddot{\phi}.
  \end{eqnarray}
  The equation of motion, Eq.~(\ref{BR34}),  is a total derivative:
  \begin{eqnarray}
      \label{BR44} \nonumber
      \frac{d}{dt}  \left[ G \frac{d}{dt} \Psi - \left( 3H\dot{\phi}^{2} G_{X} + 3H G \right) \Psi \right] = 0, \\ \nonumber \\  
        G \frac{d}{dt} \Psi - \left( 3H\dot{\phi}^{2} G_{X} + 3HG \right) \Psi = A,
  \end{eqnarray}
   where $A$ is a constant. The restriction of the model is still necessary:
    \begin{align}
        \label{BR45}
        G\frac{1}{a^{3}} \frac{d}{dt} \left( a^{3} \dot{\phi} \right) = B,
    \end{align}
    with $B$ a constant. The next step is to try to solve the system together with Friedmann equation. A simple case is when $A=0$ and $G = X$ in Eq.~(\ref{BR44}).
    \begin{figure}
            \centering
            \includegraphics[scale=1]{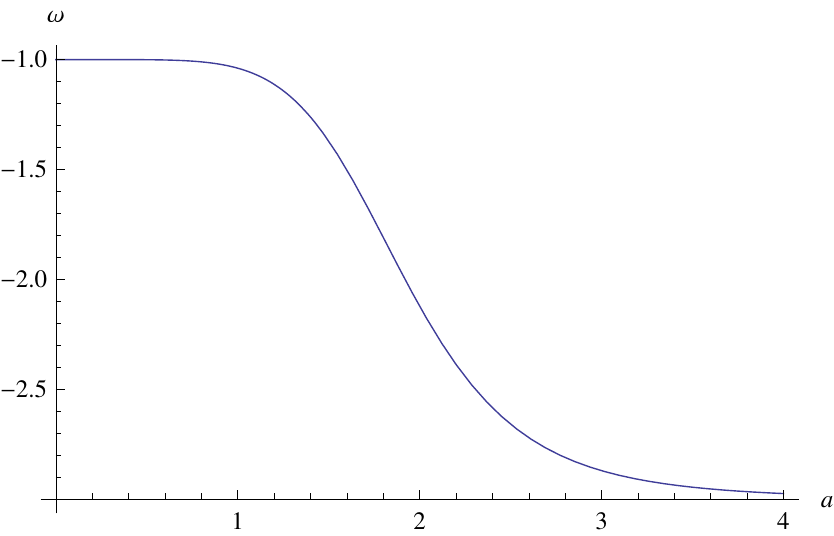}
            \caption{Equation of state parameter for the energy density and pressure given by Eqs.~(\ref{BR48}, \ref{BR49}). We take $C = \alpha =1$ and $B=-50$}
            \label{fig1}
        \end{figure}
    The equation of motion give us a solution  for $\Phi$ in terms of  the scale factor $a$:
        \begin{eqnarray}
        \label{BR46solX}
            \nonumber
            \frac{d \Psi}{dt} = 9 H \Psi, \quad \rightarrow \quad \Psi = C a^{9}, \\ \nonumber \\
            \Phi = C a^{9} - \alpha a^{3}.
        \end{eqnarray}
        We have used that $\Psi = \Phi +  \sqrt{-g}$ and $\sqrt{-g} = \alpha a$. The energy density and pressure coming from the kinetic gravity braiding model are:
        \begin{eqnarray}
            \label{BR47}
            \varepsilon = 6 H X \sqrt{2X}, \quad \mathcal{P} = -2 X \ddot{\phi}.
        \end{eqnarray}
        If one uses these expressions, together with the restriction (\ref{BR45}), in Eqs.~(\ref{BR35}, \ref{BR36}) we obtain the total energy density and pressure:
        \begin{eqnarray}
            \label{BR48}
            \varepsilon_{T} &=& B \frac{C}{\alpha} a^{6} - B, \\ \nonumber \\ 
            \label{BR49}
            \mathcal{P}_{T} &=& -3B \frac{C}{\alpha} a^{6} + B.
        \end{eqnarray}
        The equation of state parameter is defined by $\omega =
        \mathcal{P}_T / \varepsilon_T$. The figure \ref{fig1} shows the
        parameter $\omega$ as function of the scale factor $a$ for
        some particular values of the free parameters. It starts as a
        cosmological constant when $ a= 0$ and tend to $\omega= -3$
        for large values of the scale factor. For this case the model
        has the behavior of phantom dark
        energy~\cite{Basilakos:2013vya,Maggiore:2013mea,Velten:2013qna,Ludwick:2017tox}.

%%%%%%%%%%%%%%%%%%%%%%%%%%%%%%%%%%%%%%%%%
%%%%%%%%%%%%%%%%%%%%%%%%%%%%%%%%%%%%%%%%%%%%

%%%%%%%
\section{Conclusions}
The dark energy problem is one of the most intriguing challenges in
modern cosmology. The scalar fields are one of the most usual fields
employed to model dark energy. The TMT model introduces an
additional measure which has a very interesting property that unified
in a same description the dark matter and energy for the DBI
action. In this paper we have investigated, in the context of TMT, a
general form of purely kinetic k-essence field and we have shown the
same unification. 

Furthermore, we studied the properties of the kinetic braiding models
in the TMT framework and found the general expressions for the
equations of motion. We shown that the kinetic gravity braiding
models, in the general case, can be described as an imperfect fluid
with corresponding modified expressions for the energy density, the
pressure and the heat flux that reduce to the original kinetic gravity
braiding relations when the new measure is zero.

We have shown that, independent of the kinetic gravity braiding model,
by an adequate election of the new measure it is possible to obtain
the cosmological scenarios that include stiff matter or a cosmological
constant. For the stiff matter case, and by means of a particular
value of the integration constant, exists the cosmological scenario
where the scalar field does not have any effect in the dynamics of the
Universe.

In trying to find the behavior of the fluid represented by kinetic
gravity braiding model in the TMT framework, we split the task in two
simple cases depending on the form of $G$. The first one, when $G$ is
constant, lead us to a cosmological constant scenario even when in the
original kinetic gravity braiding model the energy density and the
pressure is zero. It is very interesting to note that this behavior is
the opposite situation with respect to the stiff matter special case
where the effects of the scalar field disappear on the dynamics of the
Universe. The second case, taking $G$ as a linear function of $X$,
gave us expressions of the energy density and pressure, that for a
particular choice of the free parameters, reproduce a phantom dark
energy behavior.

Finally, we have shown that the kinetic gravity braiding models give
rise to a variety of interesting cosmological effects.  A general form
of the function $G$ opens a window to study its cosmological
consequences in more detail.

\section*{Acknowledgments}
This work was partially supported by SNI-M\'exico and CONACyT research
grant: A1-S-23238. Additionally the work of RC was partially supported
by COFAA-IPN, EDI-IPN and SIP-IPN grants: 20180741 and 20195330.

%\bibliographystyle{unsrt}
%\bibliography{TwoField11bib}

\end{document}